\newcommand{\diff}{\text{d}}
\newcommand\bras[2][]{#1[ {#2} #1\rvert}
\newcommand\kets[2][]{#1\lvert {#2} #1]}

\documentclass {article} 

\usepackage{amsmath,amsfonts,amssymb,amsthm,mathrsfs}
\usepackage{dsfont}
\usepackage{mathtools,float}
\usepackage{mathrsfs}
\usepackage{booktabs}
\usepackage{mathtools}
\usepackage[hidelinks]{hyperref}
\usepackage{tikzit}

\tikzstyle{Node}=[fill=black, draw=black, shape=circle, scale=0.3px]
\tikzstyle{coh}=[fill=white, draw=black, shape=circle, scale=0.6px, line width=1px]
\tikzstyle{coh_big}=[fill=white, draw=black, shape=circle, scale=0.6px]
\tikzstyle{coh_black}=[fill=black, draw=black, shape=circle, scale=0.6px, line width=1px]

\tikzstyle{line}=[-, fill=none, line width=1.5px]
\tikzstyle{blockline}=[-, fill=black, line width=5px]
\tikzstyle{boost}=[-, fill={rgb,255: red,128; green,128; blue,128}, draw={rgb,255: red,128; green,128; blue,128}, tikzit fill={rgb,255: red,128; green,128; blue,128}, tikzit draw={rgb,255: red,128; green,128; blue,128}, line width=5px]
\tikzstyle{specialsu2}=[-, line width=5px, fill=black, draw={rgb,255: red,0; green,0; blue,189}, tikzit fill=white, tikzit draw=black]
\tikzstyle{boxdash}=[-, line width=5px, fill=black, dash pattern=on 2pt off 2pt]
\tikzstyle{arrow}=[line width=1.1px, ->]
\tikzstyle{arrowdotted}=[line width=1.1px, ->, dash pattern=on 2pt off 1.5pt]
\tikzstyle{dashed}=[-, line width=1.5px, dash pattern=on 2pt off 1pt, fill=none]
\tikzstyle{thindash}=[-, line width=0.5px, dash pattern=on 1pt off 3pt, draw={rgb,255: red,158; green,158; blue,158}]
\tikzstyle{grayfill}=[-, fill={rgb,255: red,234; green,234; blue,234}, draw=none]
\tikzstyle{thingray}=[-, line width=0.8px, draw={rgb,255: red,158; green,158; blue,158}]
\tikzstyle{arrowgray}=[draw={rgb,255: red,158; green,158; blue,158}, ->, line width=1px]

\usetikzlibrary{shapes.geometric}
\usepackage{braket}
\renewcommand\bra[1]{{\langle{#1}|}}
\makeatletter
\renewcommand\ket[1]{%
  \@ifnextchar\bra{\k@t{#1}\!}{\k@t{#1}}%
}
\newcommand\k@t[1]{{|{#1}\rangle}}
\makeatother
\usepackage[tight]{minitoc}
\usepackage{titletoc}
\usepackage{tocloft}

\titlecontents{section}
[0pt]                                               
{}%
{\small \contentsmargin{5pt}                               
    {\bfseries \thecontentslabel}\enspace}
{\contentsmargin{0pt}\large}                        
{\small \titlerule*[.5pc]{.}\contentspage}                 
[]   
\usepackage{sectsty}

\subsubsectionfont{\normalfont\itshape}

\begin{document}

\title{Biquaternions, Majorana spinors and time-like spin-foams}
\author{J. D.\ Simão\footnote{Institute for Theoretical Physics, Friedrich-Schiller-University Jena, Helmholtzweg 4, 07743 Jena, Germany. Email: \href{j.d.simao@uni-jena.de}{j.d.simao@uni-jena.de}.}}
\date{\today}
\maketitle

\begin{abstract}
\noindent This work is developed in the context of Lorentzian spin-foams with space- and time-like boundaries. It is argued that the equations describing the semiclassical regime of the various spin-foam amplitudes admit a common biquaternionic structure. A correspondence is given between Majorana 2-spinors and time-like surfaces in Minkowski 3-space based on such complexified quaternions. A symplectic structure for Majorana spinors is constructed, with which the unitary representation theory of $\mathrm{SU}(1, 1)$ is re-derived. As the main result, we propose a symplectomorphism between Majorana spinor space (with an area constraint) and $T^*\mathrm{SU}(1, 1)$, generalizing previous studies on twisted geometries to the case of time-like 2-surfaces.
\end{abstract}

\tableofcontents
\vskip-0.2em\noindent\hrulefill

\section{Introduction}

The theoretical development of spin-foam models - putative quantum gravity theories inspired by loop quantum gravity (LQG), heuristically describing spin-network state transitions - proceeds slowly but surely. The so-called EPRL model \cite{Engle:2007wy}, derived from a classical gravitational tetrad theory with an additional topological term weighted by the Immirzi parameter $\gamma$ \cite{Holst:1995pc}, has established itself as perhaps the most promising among the newer models. Its amplitude is defined for a fixed 4-simplex triangulation of a given manifold, making the the tacit assumption that the boundary tetrahedra of each 4-simplex are space-like; a later proposal by Conrady and Hnybida (CH) \cite{Conrady:2010kc} generalized this state of affairs by allowing for boundary triangles of both space- and time-like characters. 

The semiclassical regime of EPRL-type spin-foams has been investigated in recent years \cite{Barrett:2009mw, Kaminski:2017eew, Liu:2018gfc, Simao:2021qno}, and it is a general result - up to some technical obstacles in the amplitude for time-like triangles \cite{Simao:2021qno} - that, under certain circumstances, the proposed amplitude asymptotically recovers the Lorentzian Regge action  (i.e. a discretized version of the Einstein-Hilbert action). In a nutshell, this is achieved by resorting to the coherent representation of the vertex (or 4-simplex) amplitude \cite{Livine:2007vk}, schematically given by an integral
\begin{equation}
\label{vertex}
A_v=\int_{\text{SL}(2,\mathbb{C})}\prod_{a=1}^5 \diff g_a \prod_{\substack{a,b=1\\a<b}}^5 \int_{\mathbb{C}P} \omega(z_{ab})  \Omega_{ab}(z_{ab},g_a,g_b) 
\end{equation}
over 1) five copies $g_a\in \mathrm{SL}(2,\mathbb{C})$ (one for each boundary tetrahedron) of the double cover of the isometry group of $\mathbb{R}^{1,3}$, and 2) ten spinors $z_{ab} \in \mathbb{C}P$ (one for each boundary triangle) with a given measure $\omega$ (see e.g. \cite{Barrett:2009mw}). The form of the integrand $\Omega_{ab}$ depends on the particular type of vertex amplitude under consideration, and especially on the causal character of the boundary. One then performs a stationary phase approximation of $A_v$, and certain structures involving the boundary data arise which can be put into correspondence with the elements of the sphere $S^2$, the time-like hyperboloids $H^\pm$ and the space-like hyperboloid $H^{\mathrm{sl}}$ (the transitive spaces of the Lorentz group). This is enough to reconstruct tetrahedra in $\mathbb{R}^3$ and $\mathbb{R}^{1,2}$, and ultimately recover a Lorentzian 4-simplex with its associated Regge action. 

Parallel to these developments, much work has been done on the subject of spinor states in loop quantum gravity. The spinor parametrization of the LQG phase space goes by the name of ``twisted geometries'' \cite{Freidel:2010aq, Freidel:2010bw, Livine:2011gp}, hinging on the identification between $T^*\mathrm{SU}(2)$ and two copies of $\mathbb{C}^2$ spinors, together with an appropriate symplectic structure and constraint \cite{Freidel:2010bw}. This characterization allows for a geometrical interpretation of spinorial states in terms of elements of $S^2$, and thus as space-like triangles in space-like tetrahedra. A generalization to space- and time-like triangles of time-like tetrahedra and the bundle $T^*\mathrm{SU}(1,1)$ has also been proposed in the beautiful paper \cite{Rennert:2016rfp}.

Spinors thus figure in the spin-foam lore via two complementary avenues: one through the more constructive, kinematical characterization of twisted geometries, and one through the more deductive, dynamical description of the vertex amplitude. Both approaches ought to converge, and most importantly one should expect them to inform each other. 
 
\newpage 
The goal of this paper is to propose new spinor states to be put into correspondence with boundary time-like surfaces; unlike those which characterize space-like ones, these states are Majorana rather than Weyl spinors. Their introduction is motivated by the EPRL-CH vertex for a 4-simplex with time-like boundary triangles, and by the identification of a common biquaternionic structure in the semiclassical analysis of the amplitude for all causal types of boundaries. Under a natural choice of symplectic structure on the space of Majorana spinors, and together with an area constraint, a symplectomorphism to the phase space of $\mathrm{SU}(1,1)$ is established. It is moreover shown that the Majorana spinors which characterize time-like surfaces furnish a spinorial representation of the continuous series of $\mathrm{SU}(1,1)$, mutatis mutandis for Weyl spinors, space-like surfaces and the discrete series of $\mathrm{SU}(1,1)$ representations.

\section{The biquaternionic algebra}
\label{sec2}

The biquaternions, or complexified quaternions $\mathbb{H}_\mathbb{C}$, are a complex non-commutative division algebra containing elements of the form
\begin{equation}
\label{biq}
  q= \alpha + i \beta_i k^i\,, \quad \alpha\,,\,\beta_i \in \mathbb{C}\,, 
\end{equation}
where $k^i$ are the three quaternionic roots of $-1$, satisfying  $k_i k_ j = -\delta_{ij} + \epsilon_{ij}^{\;\; k} k_k$. The imaginary unit is denoted by $i$, as usual. There exist two conjugation operations, one complex $\overline{\cdot}$ and one quaternionic $\cdot^c$,
\begin{equation}
\overline{q}:=\overline{\alpha}-i\overline{\beta}_ik^i\, , \quad \quad q^c:=\alpha-i \beta_i k^i\,,
\end{equation}
which commute among themselves. Note that quaternionic conjugation is distributive under an order inversion, i.e.
\begin{equation}
(q q')^c={q'}^c q^c\,.
\end{equation}
One has projections onto the scalar $S(q)$ and vector $V(q)$ (or quaternionic) parts
\begin{equation}
S(q):=\frac{q+q^c}{2}=\alpha\,, \quad \quad V(q):=\frac{q-q^c}{2}=i \beta_i k^i\,,
\end{equation}
and there exists a complex-linear inner-product
\begin{equation}
  \braket{q,q'}:=S(q q'^c)=\alpha \alpha' - \beta^i \beta'_i \,,
\end{equation} 
inducing a complex pseudo-norm 
\begin{equation}
  |q|^2=S(q q^c):=\alpha^2-\beta^i\beta_i \, \in \mathbb{C}\,.
\end{equation} 
The above properties endow the biquaternions with a complex Minkowski structure, in that there is a vector space isometry
\begin{equation}
\begin{gathered}
 \bigl( \mathbb{H}_\mathbb{C}\,, \braket{\cdot,\cdot} \bigr) \rightarrow \left(\mathbb{C}^4,\braket{\cdot,\cdot}_\eta\right)\,, \quad   \braket{u,v}_\eta=u^T \eta v\,,\\
   q \mapsto \alpha \hat{e}_0 + \beta_i  \hat{e}^i\,.
\end{gathered}
\end{equation}
Of course, such complex quaternions offer more degrees of freedom than are necessary to describe real Minkowski space. To remedy this we can introduce a constraint, e.g. 
\begin{equation}
  q=\overline{q}^c \; \Leftrightarrow\, \alpha, \beta_i \in \mathbb{R}\,,
\end{equation}
such that the above map restricts to genuinely real Minkowski vectors. Finally, it is straightforward to see that the maps 
\begin{equation}
  L_x\,, |x|^2=1:  \; q \mapsto x q \overline{x}^c
\end{equation}
leave the norm of $q$ invariant, take real Minkowski vectors to real Minkowski vectors, and satisfy the composition law $L_{xy}=L_x L_y$, thus corresponding to Lorentz transformations. 

Since left- and right-multiplication are linear, one might hope that the biquaternionic algebra admits a realization as a matrix algebra. This is indeed the case, and the biquaternions can be realized in terms of the standard Pauli matrices $\sigma_i$ through the mapping
\begin{equation}
\begin{gathered}
\mathbb{H}_{\mathbb{C}} \rightarrow \mathbb{C}^{2\times 2} \\
  \left(1, k_i\right) \mapsto \left(\mathds{1}, -i\sigma_i \right)\,;
\end{gathered}
\end{equation}
a general biquaternion then takes the form
\begin{equation}
 q= \alpha+ i\beta_i k^i \mapsto X_q = \begin{pmatrix} \alpha + \beta_3 & \beta_1 -i \beta_2 \\ \beta_1 + i\beta_2 & \alpha-\beta_3 \end{pmatrix}\,.
\end{equation}
It is useful to extend the biquaternionic operations defined up to now to this matrix representation. Define to that end the quaternionic structure\footnote{We call $Q$ a quaternionic structure in analogy with the notion of complex structure, which
is a linear map on a real even-dimensional vector space squaring to $-1$. $Q$ is one among three anti-linear maps on $\mathbb{C}^2$ which single out an identification with $\mathbb{H}$. In the spin-foam literature this map is usually denoted $J$, e.g. in \cite{Barrett:2009mw}.}
\begin{equation}
\label{qstrut}
  \begin{gathered}
  Q: \; \mathbb{C}^2 \rightarrow \mathbb{C}^2 \\
  z \mapsto -i \sigma_2 \cdot \overline{z}\,,
  \end{gathered}
\end{equation}
which is an antilinear map such that $Q^2=-1$, acting on $g \in \mathrm{GL}(2,\mathbb{C})$ as 
\begin{equation}
g \mapsto Q g Q^{-1}=g^{A \dagger}\,.
\end{equation}
The superscript $\cdot^A$ stands for the adjugate operation, and the hermitian conjugate $\cdot^\dagger$ will always refer to the canonical scalar product in $\mathbb{C}^2$ , i.e. $g^\dagger$ = $\overline{g}^T$. Note that $Q$ appears naturally in the context of the defining representation of $\mathrm{SU}(2)$, as it commutes with its action. With these conventions, a dictionary between biquaternions and their matrix formulation can be derived; this dictionary is succinctly represented in table \ref{tab:biq}.
\begin{table}[!ht]
\caption{\small Correspondence between biquaternions and complex matrices} 
\label{tab:biq}
\centering
\begin{minipage}[t]{0.45\linewidth}\centering \small 
\begin{tabular}{ l l }
\toprule
\multicolumn{1}{c}{$\mathbb{H}_{\mathbb{C}}$} & \multicolumn{1}{c}{$\mathbb{C}^{2\times 2}$}\\
\midrule
$ q\mapsto q^c $  & $ X_q\mapsto X_q^A\quad$ \\
$q\mapsto \overline{q}$  & $ X_q\mapsto Q X_q Q^{-1} $    \\
$|q|^2=S(qq^c)$ & $|X_q|^2=\det X_q $\\
\bottomrule
\end{tabular}
\end{minipage}\hfill%
\begin{minipage}[t]{0.55\linewidth}\centering \small 
\begin{tabular}{ l l }
\toprule
\multicolumn{1}{c}{$\mathbb{H}_{\mathbb{C}}$} & \multicolumn{1}{c}{$\mathbb{C}^{2\times 2}$}\\
\midrule
$ S(q)=\frac{q+q^c}{2} $  & $ S\left(X_q\right)= \frac{\mathds{1}}{2}\mathrm{Tr} X_q $ \\
 $ V(q)=\frac{q-q^c}{2}  $  & $V \left(X_q\right)= \frac{\sigma^i}{2}\mathrm{Tr}\left[\sigma_i X_q \right]  $    \\
$\braket{q,q'}=S(q q'^c) $  & $ \braket{X_q, X_{q'}}=\frac{1}{2}\mathrm{Tr}\left[X_q X_{q'}^A \right] $    \\
\bottomrule
\end{tabular}
\end{minipage}
\end{table}

\section{Closure conditions in spin-foams}

As claimed in the introduction, the semiclassical analysis of EPRL-type models \cite{Barrett:2009mw, Kaminski:2017eew, Liu:2018gfc, Simao:2021qno} based on equation \eqref{vertex} generally yields boundary-dependent objects which admit an interpretation in terms of points in $S^2$, $H^\pm$ and $H^\mathrm{sl}$. For the readers' convenience, these objects are here recalled. 

Consider first the $\mathbb{C}^2$ pairings
\begin{equation}
\braket{u|v}:=u^\dagger v\,, \quad [u|v]:=u^\dagger \sigma_3 v\,,
\end{equation}
which are invariant under the defining action of $\mathrm{SU}(2)$ and $\mathrm{SU}(1,1)$ respectively. We accept the convention that $\ket{v}=\kets{v}$ and $\bras{v}=\bra{v}\sigma_3$, and define
\begin{equation}
\ket{+}:=\begin{pmatrix} 1 \\ 0 \end{pmatrix}\,, \quad \ket{-}:=\begin{pmatrix} 0 \\ 1 \end{pmatrix}\,, \quad   \ket{l^\pm}:=\frac{1}{\sqrt{2}} (\ket{+} \pm \ket{-})\,, \quad \ket{z}:=\begin{pmatrix} z_1 \\ z_2 \end{pmatrix}\,,
\end{equation}
further introducing the tuple
\begin{equation}
\label{varsigma1}
\varsigma:=(\sigma_3, i\sigma_2, -i\sigma_1)=\sigma_3(\mathds{1},\sigma_1,\sigma_2)\,.
\end{equation}
Let $\bar a$ denote a distinguished tetrahedron in the vertex amplitude, and $b$ a neighboring one. The integral \eqref{vertex} involves a choice of boundary data for each interface triangle $\bar a b$, composed of pairs $(j_{\bar a b}, h_{\bar a b})$ of spins\footnote{The representation theory of $\mathrm{SU}(1,1)$, along with the relevant conventions and notation, is discussed in section \ref{su11reps}.} and group elements as follows:
\begin{enumerate}
\item if $\bar a b$ is a space-like triangle of a space-like tetrahedron, then $h_{\bar a b} \in \mathrm{SU}(2)$ and $j_{\bar a b}\in \frac{\mathbb{N}}{2}$ labels a unitary irred. rep. of $\mathrm{SU}(2)$;
\item if $\bar a b$ is a space-like triangle of a time-like tetrahedron, then $h_{\bar a b} \in \mathrm{SU}(1,1)$ and $j_{\bar a b}\in -\frac{\mathbb{N}}{2}$ labels a unitary irred. rep. of the positive ($t_{\bar a b}:=+$) or negative ($t_{\bar a b}:=-$) discrete series of $\mathrm{SU}(1,1)$;
\item if $\bar a b$ is a time-like triangle of a time-like tetrahedron, then $h_{\bar a b} \in \mathrm{SU}(1,1)$ and $j_{\bar a b}\in \mathbb{R}^+$ labels a unitary irred. rep. of the continuous series\footnote{Later on the notation $j$ will stand for the complex parameter $j=-1/2+i s$ of the continuous series, with $s$ the real continuous parameter. Here $j$ is used as a stand-in for $s$ in order to simplify the exposition.} of $\mathrm{SU}(1,1)$.
\end{enumerate}
A \textit{necessary} condition \cite{Barrett:2009mw, Kaminski:2017eew, Liu:2018gfc, Simao:2021qno} for \eqref{vertex} to have critical points is for the following \textit{closure conditions} to hold: if $\bar a$ is a assumed to be a space-like tetrahedron, then
\begin{equation}
\label{closure1}
\sum_b j_{\bar a b} \braket{+_{\bar a b}| \sigma_i +_{\bar a b}}=0\,, \quad \ket{+_{\bar a b}}:=h_{\bar a b}\ket{+}\,;
\end{equation}
if $\bar a$ is time-like, containing both space- (s.l) and time-like (t.l.) triangles, then
\begin{equation}
\label{closure2}
\sum_{\substack{b\\ \bar a b \;\mathrm{s.l.}}}j_{\bar a b} t_{\bar a b}[t_{\bar a b}|\varsigma_i  t_{\bar a b}] + i \sum_{\substack{b\\ \bar a b \;\mathrm{t.l.}}}{j_{\bar a b}}[l^-_{\bar a b}|\varsigma_i l^+_{\bar a b}]=0\,, \quad |l^\pm_{\bar a b}]:=h_{\bar a b}|l^\pm]\,.
\end{equation}

The geometrical interpretation of the spin-foam semiclassical regime hinges on the identification of the terms $\braket{+_{\bar a b}| \sigma_i +_{\bar a b}}$, $[t_{\bar a b}|\varsigma_i  t_{\bar a b}]$ and $[l^-_{\bar a b}|\varsigma_i l^+_{\bar a b}]$ with 3-vectors $v_i$ as elements of $S^2, H^\pm$ and $H^{\mathrm{sl}}$, respectively. Minkowski's theorem on Lorentzian convex polyhedra \cite{Simao:2021qno} can then be used to uniquely reconstruct geometrical tetrahedra with triangle areas $j_{\bar ab}$. This correspondence, which may be evaluated explicitly, can further be clarified by resorting to the machinery of biquaternions discussed in section \ref{sec2}. This we do in the following.   

\section{Spinors and homogeneous spaces}

Throughout this section the letter $q$ will be used to denote both a biquaternion proper as well as - by abuse of notation - its $\mathbb{C}^{2\times 2}$ matrix representation. We endeavor to associate a general spinor $\ket{z}\in \mathbb{C}^{2}$ to the various $\mathbb{R}^{1,3}$ subspaces $S^2, H^\pm$ and $H^{\mathrm{sl}}$. This will be done by first mapping the spinor to a biquaternion, which then admits a geometrical interpretation in Minkowski space as per the discussion of section \ref{sec2}. 

To begin, consider a general biquaternion $q\in \mathbb{C}^{2\times 2}$ - argued above to be in correspondence with complex Minkowski vectors - and reduce its dimensionality by requiring $\det q=0$. Table \ref{tab:biq} shows this implies $|q|^2=0$, or in terms of biquaternionic components
\begin{equation}
\label{comps}
\beta_1^2+\beta_2^2+\beta_3^2=\alpha^2\,,
\end{equation}
i.e. we restrict to \textit{null} biquaternions. The three subspaces $S^2, H^\pm, H^{\mathrm{sl}}$ are now treated individually\footnote{The case for $S^2$ amounts to the well-known Bloch-sphere construction.}.

\subsection{The space-like sphere $S^2$}

Note from the components \eqref{comps} that the defining equation for the sphere $S^2:=\{(x,y,z) \in \mathbb{R}^3 \;| \; x^2+y^2+z^2=1 \}$ is recovered if all $\beta_i,\alpha$ are required to be real. As pointed out in section \ref{sec2}, and referring to table \ref{tab:biq}, this can be achieved through the constraint $q=q^\dagger$. Clearly $q_z=\ket{z}\bra{z}$ satisfies both $\det q_z=0$ and $q_z=q_z^\dagger$, and thus one can model space-like vectors of $\mathbb{R}^3$ through the sequence of bijections (note that there is a phase redundancy) 
\begin{equation}
  \begin{gathered} 
 \mathbb{C}^2 / \mathrm{U}(1)\; \rightarrow \; \mathbb{C}^{2\times 2}_{\substack{\det q=0\\ q=q^\dagger}} \; \rightarrow \; \left(\mathbb{R}^3, \eta_{(3)} \right) \\
 \ket{z} \; \mapsto \; q_z=\ket{z}\bra{z}\;  \mapsto\; v_z^i=\frac{1}{2} \mathrm{Tr}\left[\sigma^i q_z \right]\,.
  \end{gathered}
\end{equation}
Here $\eta_{(3)}=\mathrm{diag}(-1,-1,-1)$ denotes the induced metric on $\mathbb{R}^3\subset \mathbb{R}^{1,3}$ defined by
\begin{equation}
\begin{gathered}
  \eta_{(3)}(v_z,v_w):=\braket{V( q_z ), V(q_w)}=-{v_z}^i {v_w}_i\,, 
\end{gathered}
\end{equation} 
and equation \eqref{comps} guarantees that $|v_z|^2=-\left(\frac{1}{2}\mathrm{Tr} q_z\right)^2$. Under a suitable rescaling, unit-norm spinors can then be put in correspondence with the sphere as
\begin{equation}
 \ket{z}\;\, \mathrm{s.t.} \; \braket{z|z}^2=1  \; \mapsto \; v_z^i=\braket{z|\sigma^i|z} \in S^2\,,
\end{equation}
to be compared with the objects of equation \eqref{closure1}.
\subsection{The time-like two-sheeted hyperboloid $H^\pm$}
\label{spinorsl}

The unit time-like hyperboloid is defined as $H^\pm =\{(t,x,y) \in \mathbb{R}^{1,2} \;| \; t^2-x^2+y^2=1\,, \; \pm t >0  \}$. Following the same strategy, we would like to modify the above procedure in such a manner that the induced metric has signature $(+,-,-)$, and such that the resulting vectors have positive norm. This can be achieved by first permuting the components of $q$ so that the projection operation $q\mapsto V(q)$ maps to the intended subspace; right-multiplication by the third unit $k^3$ gives
\begin{equation}
\begin{gathered}
q k^3=-i \beta_3 + i \left(\beta_2 k_1 -\beta_1 k_2 -i \alpha k_3 \right)\,, \\
  |V(q k^3)|^2 = \alpha^2-\beta_1^2-\beta_2^2\,, 
  \end{gathered}
\end{equation}
as desired. The construction then follows as before: we impose $\det q =0$ and $q=q^\dagger$, but we right-rotate the quaternion $q \mapsto q(-i \sigma_3)$. The sequence of bijections
\begin{equation}
\label{componentssl}
  \begin{gathered} 
 \mathbb{C}^2/\mathrm{U}(1)\; \rightarrow \; \mathbb{C}^{2\times 2}_{\substack{\det q=0\\ \sigma_3 q\sigma_3=-q^\dagger}} \; \rightarrow \; \left(\mathbb{R}^{1,2}_{|\cdot|^2\geq 0,\, v^1\geq 1}, \eta_{(1,2)} \right) \\
 \ket{z} \; \mapsto \; q_z=-i \ket{z}\bra{z} \sigma_3\;  \mapsto\; v_z^i=\frac{i}{2} \mathrm{Tr}\left[\varsigma^i q_z \right]
  \end{gathered}
\end{equation}
yields the subspace of future-pointing\footnote{The map only covers the future-pointing hyperboloid, as it can be checked that $v_z^1\geq 1$. The lower hyperboloid can be obtained by taking the symmetric of the components in equation \eqref{componentssl}.} time-like vectors in $\mathbb{R}^{1,2}\subset \mathbb{R}^{1,3}$. The particular form of $\varsigma$, introduced in equation \eqref{varsigma1}, was chosen in order to get the right vector components\footnote{$\varsigma$ can also be seen as a vector of $\mathrm{SU}(1,1)$ generators in the defining representation; we will later make use of this fact.} $v_z=(\alpha, \beta_1,\beta_2)$.
The induced metric takes the form 
\begin{equation}
  \eta_{(1,2)}(v_z,v_w):=\braket{V( q_z ), V(q_w)}=v_z^i v_w^j \eta_{(1,2)ij} \,, 
\end{equation}
with $\eta_{(1,2)}=\mathrm{diag}(1,-1,-1)$, and the norm is constrained by equation \eqref{comps} to satisfy
\begin{equation}
\label{n_pos}
  |v_z|^2=\beta^2_3=\left(\frac{i}{2} \mathrm{Tr} q_z \right)^2 \geq 0\,,
\end{equation}
since $\beta_3$ was chosen to be real; all vectors so constructed are indeed time-like. The Minkowski metric is also naturally associated to $\varsigma$, since
\begin{equation}
\label{varsigma}
  \varsigma^i \varsigma^j = \eta_{(1,2)}^{ij} -i \epsilon^{ijk}\eta_{(1,2)kl} \varsigma^l\,.
\end{equation}
Finally, the mapping from spinors to the two-sheeted hyperboloid reads
\begin{equation}
\label{smatch}
 \ket{z}\;\, \mathrm{s.t.} \; [z|z]^2=1  \; \mapsto \; v_z^i=\pm [z| \varsigma^i |z] \in H^\pm\,.
\end{equation}
Note that this construction has the right symmetry properties, since the vector norm is invariant under the natural action $\ket{z} \mapsto g\ket{z}$ of $g\in \mathrm{SU}(1,1)$ on $\mathbb{C}^2$, which is (the double-cover of) the isometry group of $\mathbb{R}^{1,2}$. The very same vector components $v_z^i=\pm [z| \varsigma^i |z]$ appear in the closure relation \eqref{closure2}.

\subsection{The space-like one-sheeted hyperboloid $H^{\mathrm{sl}}$}
\label{spinortl}

The unit space-like hyperboloid is given by the set $H^{\mathrm{sl}} =\{(t,x,y) \in \mathbb{R}^{1,2} \;| \; t^2-x^2+y^2=-1 \}$. As equation \eqref{comps} shows, the induced vector norm $\alpha^2-\beta_1^2-\beta_2^2$ will be positive as long as $\beta_3$ remains real; it must thus be made purely imaginary, and this requires amending the hermiticity condition. One should consider instead
\begin{equation}
  k^3 q k^3 = - \overline{q} \quad \Leftrightarrow \quad \alpha,\beta_1,\beta_2 \in \mathbb{R}\,, \; \beta_3 \in i \mathbb{R}\,,
\end{equation}
which in the matrix formulation reads $\sigma_3 q \sigma_3 = Q q Q^{-1}$. The real structure $R^2=1$ associated with $\mathrm{SU}(1,1)$ 
\begin{equation}
  \begin{gathered}
  R: \; \mathbb{C}^2 \rightarrow \mathbb{C}^2 \\
  \ket{z} \mapsto \sigma_1 \ket{\overline{z}}\
  \end{gathered}
\end{equation}
naturally emerges, and the constraint has the equivalent form $q=RqR^{-1}$. Note that $R$ commutes with $g\in \mathrm{SU}(1,1)$, i.e. $Rg=gR$. 

The solutions to $\det q=0,\,q=Rq R^ {-1}$ are less immediate than in the previous cases. Because $q$ is singular, it can be written as the exterior product of two elements of $\mathbb{C}^2$, i.e $q=\ket{x}\bra{y}$. The second constraint then implies $R \ket{x} = \lambda_y \ket{x}$ and $R \ket{y} = \lambda_x \ket{y}$ with $\lambda_x \overline{\lambda}_y=1$ some complex numbers depending on $x$ and $y$, respectively. Up to multiplicative factors we are thus interested in eigenstates of the real structure, and these are given in terms of a general spinor as
\begin{equation}
\ket{z^\pm}=\frac{1}{\sqrt{2}} \left(\ket{z}\pm R\ket{z} \right)\,, \\
\end{equation}
with eigenvalues $R \ket{z^\pm} = \pm \ket{z^\pm}$. Hence we might propose two classes of solutions to the constraints,
\begin{equation}
  \begin{cases}
  q = \ket{z^\pm}\bra{z^\pm}\,, \\
  q= \alpha \ket{z^\mp} \bra{z^\pm}\,,
  \end{cases}
\end{equation}
where $\alpha$ is some linear map anticommuting with the real structure. 
One can however quickly convince themselves that the first alternative is over-constrained, leading exclusively to null vectors\footnote{Unsurprisingly so, since it is hermitian, thus satisfying the constraints associated to both space- and time-like vectors.} in $\mathbb{R}^{1,2}$. We are left with the second option, and the choice $\alpha=\sigma_3$ will prove to be particularly useful in simplifying calculations. 

Before proceeding we address the cumbersomeness of the states $\ket{z^\pm}$ by performing a change of variables 
\begin{equation}
\label{M1}
\sigma_3 \ket{z^-} = \begin{pmatrix} z_1 - \overline{z}_2 \\ \overline{z}_1-z_2 \end{pmatrix} \mapsto \ket{z^1}:= \begin{pmatrix} z_1\\  \overline{z}_1  \end{pmatrix}\,,
\end{equation}
\begin{equation}
\label{M2}
\ket{z^+} = \begin{pmatrix} z_1 + \overline{z}_2 \\ z_2 + \overline{z}_1 \end{pmatrix} \mapsto \ket{z^2}:= \begin{pmatrix} z_2\\  \overline{z}_2  \end{pmatrix}\,.
\end{equation}
This justifies having picked $\alpha=\sigma_3$, which was done so that $\ket{z^1}$ and $\ket{z^2}$ have the same qualitative structure. In this manner the two complex degrees of freedom of a Weyl spinor have been split across two Majorana spinors - i.e. spinors which are invariant under the real structure, known as the charge conjugation operator in QFT - whose components are constrained to be related by complex conjugation. This signifies a substantial departure from the previous $S^2$ and $H^\pm$ cases, which could be described by a single Weyl spinor. Note moreover that, since the real structure commutes with $\mathrm{SU}(1,1)$, the natural action $g \triangleright \ket{z^{1,2}}= g \ket{z^{1,2}}$ takes Majorana spinors to Majorana spinors.  

We can now state the sequence of bijections yielding space-like vectors. Considering again the right-product with the third quaternionic unit, they read
\begin{equation}
  \begin{gathered} 
 (\mathbb{C}_{\mathrm{Majorana}}\oplus \mathbb{C}_{\mathrm{Majorana}})/\mathrm{U}(1)\; \rightarrow \; \mathbb{C}^{2\times 2}_{\substack{\det q=0\\ R q R^{-1}= q}} \; \rightarrow \; \left(\mathbb{R}^{1,2}_{|\cdot|^2 \leq 0}, \eta_{(1,2)} \right) \\
 \left(\ket{z^1}, \, \ket{z^2}\right) \; \mapsto \; q_z= -i \ket{z^1}\bra{z^2}\sigma_3 \;  \mapsto\;  v_z^i=\frac{i}{2}\mathrm{Tr}\left[\varsigma^i q_z \right] \,.
  \end{gathered}
\end{equation}
with the vector norm given by 
\begin{equation}
  |v_z|^2=\beta^2_3=\left(\frac{i}{2} \mathrm{Tr} q_z \right)^2 \leq 0\,,
\end{equation}
which is again invariant under the $\mathrm{SU}(1,1)$ action, as intended. The one-sheeted hyperboloid can finally be written in terms of two Majorana spinors as
\begin{equation}
\label{tmatch}
 \ket{z^1},\, \ket{z^2} \;\, \mathrm{s.t.} \; \braket{z^2|\sigma_3|z^1}^2=-1  \; \mapsto \; v_z^i =\braket{z^2|\sigma_3 \varsigma^i|z^1} \in H^{\mathrm{s.l.}}\,.
\end{equation} 
Again, note that $v_z^i =\braket{z^2|\sigma_3 \varsigma^i|z^1}$ makes an appearance in equation \eqref{closure2}.

{\centering \noindent\rule{2cm}{0.3pt} \\~\\}

\noindent The discussion above shows that the geometrical spaces of interest for EPRL-type spin-foams can be modeled on Weyl and Majorana spinors. The usage of Majorana rather than Weyl in the context of  $H^{\mathrm{sl}}$ departs from the orthodox prescription of twisted geometries, and in particular from Rennert's proposal \cite{Rennert:2016rfp}; it is therefore still necessary to show that such spinors can be used to characterize the phase space $T^* \mathrm{SU}(1,1)$. As an intermediate step to that end, we will first argue that the Weyl and Majorana spinors associated with $H^\pm$ and $H^{\mathrm{sl}}$ lead to a spinorial realization of the discrete and continuous series (respectively) of the unitary representations of $\mathrm{SU}(1,1)$, thus obtaining an alternative constructive derivation of the Conrady-Hnybida prescription \cite{Conrady:2010kc}.

\section{Spinorial realizations of $\mathrm{SU}(1,1)$ representations}
\label{su11reps}

It was proven in \cite{Livine:2011gp, Livine:2011vk} that the parametrization of $\mathrm{SU}(2)$ elements in terms of Weyl spinors could be used to derive the Bargmann-Fock measure \cite{Woit:2017vqo} on holomorphic functions from the Haar measure; the former corresponds to the inner product on the supporting Hilbert spaces of $\mathrm{SU}(2)$ unitary representations. It stands to reason that a similar result ought to be achievable for $\mathrm{SU}(1,1)$. 

In this section the unitary representations of $\mathrm{SU}(1,1)$ are constructed from Weyl and Majorana spinors $\ket{z}$ and $\ket{z^{1,2}}$, respectively. The strategy for both cases - to be treated separately - will be to first postulate an appropriate symplectic structure on the relevant space, leading to a spinor realization of the $\mathfrak{su}(1,1)$ algebra, and thus of the group itself.   

\subsection{The discrete series from $H^\pm$ Weyl spinors}

\subsubsection{Poisson structure}

Recall from section \ref{spinorsl} that the two-sheeted hyperboloid $H^\pm$ can be modeled on a Weyl spinor $\ket{z}\in \mathbb{C}^2$ as per equation \eqref{smatch}. In the spirit of Bargmann-Fock quantization\footnote{The well-known Bargmann-Fock quantization of an harmonic oscillator starts by picking a complex structure in order to complexify the phase space. Complex linear combinations of the canonical variables $(q,p)$ then lead to creation and annihilation operators. Since here we already start with a complex vector space, we consider simply a direct sum with its complex conjugate.} we consider the larger space $\mathbb{C}^4\simeq \mathbb{C}^2 \oplus \overline{\mathbb{C}}^2$, together with a distinguished basis $\{\partial_1, \partial_2\}\oplus \{\overline{\partial}_1, \overline{\partial}_2\}$. A general element in $\mathbb{C}^4$ can be written as $u=u^+ + u^-$, where $u^+ \in \mathbb{C}^2$ and $u^- \in \overline{\mathbb{C}}^2$. Complex conjugation takes $u^+ \in \mathbb{C}^2$ to $\overline{u}^+ \in \overline{\mathbb{C}}^2$, and vice-versa.  The discussion of section \ref{spinorsl} suggests that $\mathbb{C}^2$ should be equipped with the indefinite hermitian form
\begin{equation}
\begin{gathered}
  \braket{\cdot,\cdot}_{\mathbb{C}^2}: \; \mathbb{C}^2\times \mathbb{C}^2 \rightarrow \mathbb{C} \\
  (u^+,v^+) \mapsto \braket{u^+,v^+}_{\mathbb{C}^2}=u^{+\dagger} \sigma_3 v^+\,,
\end{gathered}
\end{equation}
which is invariant under the natural action of $\mathrm{SU}(1,1)$ on $\mathbb{C}^2$. There is a standard construction for assigning a symplectic structure given such a pairing \cite{Woit:2017vqo}: one defines
\begin{align*}
  & i \omega( u^-, v^+)=  \braket{\overline{u}^-,v^+}_{\mathbb{C}^2}\,, \\
  &  \omega(u^+, v^+)= \, \omega(u^-, v^-)=0\,,
\end{align*}
and the requirement of antisymmetry induces another hermitian form in $\overline{\mathbb{C}}^2$ as $\braket{u^-,v^-}_{\overline{\mathbb{C}}^2}=- u^{+\dagger} \sigma_3 v^+$. The symplectic form thus reads
\begin{equation}
\label{symplecticsl}
  \omega=i \sigma_3^{ij} \, \diff z_i \wedge \diff \overline{z}_j \,,
\end{equation}
differing from the standard symplectic structure \cite{Speziale:2012nu} on $\mathbb{C}^2 \oplus \overline{\mathbb{C}}^2$ by the inclusion of the third Pauli matrix. It induces Poisson brackets on coordinate functions 
\begin{equation}
\label{pissonmajorana}
  \{\overline{z_i}, z_j\}=i {\sigma_3}_{ij}\,,
\end{equation}
all other brackets vanishing. 

\subsubsection{Lie algebra realization}

Consider the geometric vector components of equation \eqref{componentssl}, now denoted by $v^i=\frac{1}{2} \braket{z|\sigma_3 \varsigma^i |z}$,
\begin{equation}
  v^1=\frac{1}{2}(|z_1|^2+|z_2|^2)\,, \quad v^2=\frac{1}{2}(z_1 \overline{z}_2 + \overline{z}_1 z_2)\,, \quad v^3=\frac{i}{2}(z_1 \overline{z}_2-\overline{z}_1 z_2)\,,
\end{equation}
with vector norm
\begin{align}
|v|^2&=(v^1)^2-(v_2)^2-(v_3)^2 \nonumber \\
&=\frac{1}{4}(|z_1|^2-|z_2|^2)^2 \geq 0\,.
\end{align}
A straightforward calculation employing the identity \eqref{varsigma} shows that
\begin{equation}
\label{algsl}
  \{v^i, v^j\}=-\epsilon^{ijk}\eta_{kl} v^l\,,
\end{equation}
where for the rest of this paper $\eta:=\eta_{(1,2)}$. The functions $v^i$ induce Hamiltonian vector fields $X^i$ through the symplectic form $\omega(X^i, \cdot)=\diff v^i(\cdot)$, which explicitly read
\begin{equation}
  X^1=-\frac{i}{2}\left(z_1 \partial_1 - z_2 \partial_2 +  \overline{z}_2 \overline{\partial}_2 - \overline{z}_1 \overline{\partial}_1 \right)\,,
\end{equation}
\begin{equation}
  X^2=-\frac{i}{2}\left(z_2 \partial_1 - z_1 \partial_2 + \overline{z}_1 \overline{\partial}_2 - \overline{z}_2 \overline{\partial}_1 \right)\,,
\end{equation}
\begin{equation}
  X^3=-\frac{1}{2}\left(z_2 \partial_1+z_1 \partial_2 +\overline{z}_2 \overline{\partial}_1 + \overline{z}_1 \overline{\partial}_2\right)\,.
\end{equation}
and which satisfy the Lie bracket relations $[X^i, X^j]= \epsilon^{ijk}\eta_{kl} X^l$. The reader may recognize in this equation the Lie brackets of $\mathfrak{su}(1,1)$, suggesting the vector fields can be thought of as its generators; the algebra has however too many degrees of freedom. Recalling the orthogonal decomposition $\mathbb{C}^2 \oplus \overline{\mathbb{C}}^2$ - which was itself used in the construction of the symplectic structure -, we therefore choose to project the vectors onto the first factor $\mathbb{C}^2$. Under the identification $( X^1, X^2,  X^3) \mapsto (iL^3, iK^1, iK^2)$ (which includes imaginary units to match the physics literature \cite{Conrady:2010sx}), one gets the vector fields
\begin{equation}
\label{gensl}
  L^3=\frac{1}{2}\left(z_2 \partial_2-z_1 \partial_1 \right)\,, \quad  K^1=\frac{1}{2}\left(z_1 \partial_2-z_2 \partial_1\right)\,, \quad K^2=\frac{i}{2}\left(z_1 \partial_2+z_2 \partial_1\right)\,, 
\end{equation}
together with the algebra
\begin{equation}
  [L^3,K^1]=iK^2\,, \quad [L^3,K^2]=-iK^1\,, \quad [K^1,K^2]=-i L^3\,.
\end{equation}
The vector fields so constructed thus constitute a spinorial representation of the Lie algebra. In this realization the Casimir element $Q$ takes the form
\begin{align}
\label{casimirsl}
  Q&=(L^3)^2-(K^1)^2-(K^2)^2 \nonumber \\
  &=\frac{1}{4} \left[ 2(z_1 \partial_2 + z_2 \partial_2) + (z_1 \partial_2 + z_2 \partial_2)^2 \right] \,,
\end{align}
and one can define the objects
\begin{equation}
  L^\pm=K^1 \pm i K^2\,, \quad L^+=-z_2 \partial_1\,, \quad L^-=z_1 \partial_2\,,
\end{equation}
which will soon be shown to act as creation and annihilation operators, respectively.

\subsubsection{Group representations}

The $\mathfrak{su}(1,1)$ generators obtained above are all holomorphic derivations depending on coordinate functions $z_1$ and $z_2$. This suggests seeking representations of $\mathrm{SU}(1,1)$ on the ring of formal power series $\mathbb{C}[[z_1,z_2]]$. Indeed, the algebra representation \eqref{gensl} corresponds to the usual group representation in terms of operators on a function space
\begin{equation}
  \begin{gathered}
  D: \mathrm{SU}(1,1) \rightarrow \mathcal{O}\left(\mathbb{C}[[z_1,z_2]]\right) \\
  D(g) f(z_1, z_2) =  f ( g^{-1}\triangleright  (z_1, z_2))\,, \quad g\triangleright  (z_1, z_2)= g\begin{pmatrix}
    z_1 \\ z_2 \end{pmatrix}\,,
  \end{gathered}
\end{equation} 
as can be retroactively checked by setting $(L^3, K^1, K^2)=\frac{1}{2}(\varsigma_1, \varsigma_2, \varsigma_3)$ in the defining representation, i.e.
\begin{equation}
L^3=D'(\sigma_3/2)\,, \quad K^1=D'(i\sigma_2/2)\,, \quad K^2=D'(-i\sigma_1/2)\,, 
\end{equation}
where $D'(X)$ stands for the induced Lie algebra representation. For example, considering $K^1$ one finds
\begin{align}
  D'\left(\frac{i\sigma_2}{2}\right)f(z_1,z_2)&:=-i \frac{\diff}{\diff t}|_{t=0} D\left( e^{i \frac{i\sigma_2 }{2}t}\right) f(z_1,z_2) \nonumber \\
  &=\frac{1}{2}\left(z_1 \frac{\partial f}{\partial z_2}-z_2 \frac{\partial f}{\partial z_1} \right)\,,
\end{align} 
which is exactly the action of $K^1$ on $f(z_1,z_2)$ according to \eqref{gensl}.

As per Schur's lemma \cite{Woit:2017vqo}, in any irreducible representation the Casimir operator is proportional to the identity. Hence in searching for irreducible  representations it is useful to consider the eigenfunctions of the Casimir element $Q$. Clearly, any solution to 
\begin{equation}
\label{euler}
  (z_1 \partial_2 + z_2 \partial_2) f(z_1,z_2) = 2k f(z_1,z_2)\,,
\end{equation}
will also be an eigenfunction of $Q$ by virtue of equation \eqref{casimirsl}. But Euler's theorem on homogeneous functions dictates that every maximal smooth solution to such an equation is an homogeneous function of degree $2k \in \mathbb{Z}$, i.e. $f(\alpha z_1, \alpha z_2)=\alpha^{2k} f(z_1,z_2)$. The eigenfunctions of $Q$ are therefore the homogeneous series
\begin{equation}
  Q f_k(z_1,z_2) = k(k+1) f_k (z_1,z_2)\,.
\end{equation}

\subsubsection{Hermitian inner product and unitarity}

We are interested in unitary representations, and this requires finding an invariant hermitian form on $\mathbb{C}[[z_1,z_2]]$. Our strategy will be to derive such a form from the $L^2(\mathrm{SU}(1,1))$ inner product, since 1) the Haar measure is left- and right-invariant and 2) $\mathrm{SU}(1,1)$ can be parametrized in terms of the spinor $\ket{z}$. With the parametrization
\begin{equation}
\label{param}
g(z_1, z_2)=\begin{pmatrix} z_1 & z_2 \\ \overline{z}_2 & \overline{z}_1 \end{pmatrix}\,,
\end{equation}
the Haar measure takes the form $\diff g(z)=\pi^{-2} \delta(\braket{z|\sigma_3|z}-1)\, Dz_1  Dz_2$ \cite{ruhl1970lorentz}, where $D z_i:= \frac{i}{2} \diff z_i \wedge \diff \overline{z}_i$. Let $f_k,g_{k'}$ be two homogeneous functions of degree $2k,2k'$, respectively. Then
\begin{align}
  &\frac{1}{\pi^2} \int  Dz_1 Dz_2 \; \delta(\braket{z|\sigma_3|z}-1) \overline{f_k(z_1,z_2)} g_{k'}(z_1,z_2)\nonumber \\
  &= \frac{1}{2\pi^2} \int \diff \theta_1 \diff \theta_2 \diff r \, \frac{r}{(r^2-1)^2} \Theta(r-1) \overline{f}_k\left(\frac{r e^{i \theta_1}}{\sqrt{r^2-1}}, \frac{e^{i \theta_2}}{\sqrt{r^2-1}} \right) g_{k'}\left(\frac{r e^{i \theta_1}}{\sqrt{r^2-1}}, \frac{e^{i \theta_2}}{\sqrt{r^2-1}} \right) \nonumber \\
  \label{calc}
  &= \frac{i}{2\pi} \delta_{kk'} \int_{D^1} Dz\, (1-|z|^2)^{-2k-2} \overline{f}_k(1,z) g_{k'}(1,z)\,,
\end{align}
where $z_i=:\rho_i e^{i \theta_i}$, $r:= \rho_1 / \sqrt{\rho_1^2-1}$, $z^{-1}:=re^{i\theta_1}$ and $D^1$ denotes the unit disk in $\mathbb{C}$. One can then define the inner product
\begin{equation}
\label{innersl}
  \braket{f,g}:= \frac{1}{\pi} \int_{D^1} Dz\, (1-|z|^2)^{-2k-2} \overline{f}(1,z) g(1,z)\,,
\end{equation}
which is by construction invariant, i.e. $\braket{D(g)f, D(g)h}=\braket{f,g}$ and well-defined for $k\leq -1$. Following Bargmann \cite{Bargmann:1946me} we extend this inner product to the case $k=-1/2$ by setting\footnote{That \eqref{calc} glosses over the lowest spin representation is a consequence of the fact that $k=-1/2$ is absent from the Fourier series of square-integrable functions on the group, cf. \cite{ruhl1970lorentz}.}
\begin{equation}
    \braket{f,g}_{-\frac{1}{2}}:= \lim_{k \to -\frac{1}{2}}\, \frac{-2k-1}{\pi} \int_{D^1} Dz\, (1-|z|^2)^{-2k-2} \overline{f}(1,z) g(1,z)\,.
\end{equation}
The ring $\mathbb{C}[[z_1,z_2]]$ can now be restricted the Hilbert space $\mathcal{B}_k(D^1)$ of holomorphic functions on the disk.

The invariant pairing of equation \eqref{innersl} is precisely the inner product on the discrete series of unitary representations of $\mathrm{SU}(1,1)$ \cite{Bargmann:1946me}, showing the manner in which the one-sheeted hyperboloid is associated to the discrete series. The representation $D(g)f(z_1,z_2)=f(g^{-1} (z_1,z_2))$ reduces to the more common multiplier representation derived by Bargmann \cite{Bargmann:1946me} on the disk $\psi(z):=f(1,z)$ when homogeneity is taken into account:
\begin{equation}
\label{discretepos}
  D(g) \psi(z)= (\overline{\alpha} - \beta z)^{2k} \psi\left(\frac{\alpha z -\overline{\beta}}{\overline{\alpha} - \beta z} \right)\,, \quad g=\begin{pmatrix}
    \alpha & \beta \\ \overline{\beta} & \overline{\alpha}
  \end{pmatrix}\,. 
\end{equation}
One furthermore has from the above law that
\begin{equation}
  D(\mathds{1})\psi(z)=D(e^{4\pi i L^3})\psi(z)=e^{-4i\pi k}\psi(z)\,,
\end{equation}
and requiring single-valuedness restricts $k$ to be half-integer, and thus to lie in the range $k=-\frac{1}{2}-\frac{\mathbb{N}^0}{2}$. The Casimir element is therefore non-negative
\begin{equation}
  Q \sim k(k+1) \geq 0\,,
\end{equation} 
and the sign of $D(- \mathds{1})=\pm \mathds{1}$ is controlled by whether $k$ takes an integral or definite half-integral value. 

\subsubsection{The $L^3$ eigenbasis}

Consider the functions 
\begin{equation}
\label{prepold}
  f_{k,m}(z_1,z_2)=\frac{1}{\sqrt{\gamma_{k,m}}} z_1^{k-m} z_2^{k+m} \quad \left(\; \underset{D^1}{\rightarrow} \;  \psi_m(z)=\frac{1}{\sqrt{\gamma_{k,m}}} z^{k+m}\right)\,,
\end{equation}
where $\gamma_{k,m}=\frac{\Gamma(-2k-1)\Gamma(1+k+m)}{\Gamma(m-k) }$, which are clearly homogeneous of degree $2k$. They are also eigenfunctions of $L^3$, since 
\begin{equation}
  L^3 f_{k,m}=m f_{k,m}\,.
\end{equation}
The requirement that $\psi_{m}$ reduces to a polynomial on $D^1$ implies that the magnetic index takes the range $m=-k+\mathbb{N}^0$. Together with the transformation rule of equation \eqref{discretepos}, this characterizes the \textit{positive series} $\mathcal{D}^+_k$ as defined by \cite{Bargmann:1946me}. The functions $\psi_m$ are moreover orthonormal under \eqref{innersl}, as can be shown using the integral representation of the Beta function, and hence they constitute an orthonormal basis for $\mathcal{D}^+_k$. The ladder operators act as
\begin{equation}
  L^\pm f_{k,m} = \sqrt{(m\pm k\pm 1)(m\mp k)} f_{k,m\pm 1}
\end{equation}
and $L^-$ annihilates the lowest weight state with $m=-k$. 

The \textit{negative series} $\mathcal{D}^-_k$ can be obtained in a similar manner. There is an obvious outer automorphism of the Lie algebra \eqref{gensl} leaving the brackets invariant, namely the one induced by taking $(z_1, z_2) \mapsto (z_2,z_1)$. Denoting the automorphism by $P$ (for parity, to match the terminology of \cite{Lindblad:1969zz}), it reads
\begin{equation}
\label{parity}
  PL^3 P^{-1}=-L^3\,, \quad P K^1 P^{-1}= -K^1\,, \quad P K^2 P^{-1}= K^2\,, 
\end{equation}
The parity-transformed algebra can be integrated to a group representation
\begin{equation}
  D(g) f(z_1, z_2) \mapsto  f ( \overline{g}^{-1}\triangleright  (z_1, z_2))\,, 
\end{equation} 
such that 
\begin{equation}
  D(g) \psi(z)= (\alpha - \overline{\beta} z)^{2k} \psi\left(\frac{\overline{\alpha} z -\beta}{\alpha - \overline{\beta} z} \right)\,, 
\end{equation}
agreeing with the original findings of \cite{Bargmann:1946me} for $\mathcal{D}^-_k$. One can then see from equation \eqref{prepold} that under $P L^3 P^{-1}$ the magnetic index takes the values $m=k-\mathbb{N}^0$, and this is indeed the negative series. We collect both positive and negative series under the notation $\mathcal{D}^q_k$, with $q=\pm$.

The representations just constructed are known to be irreducible. A standard proof of this fact using the ladder operators can be found in \cite{combescure2012coherent}.

\subsection{The continuous series from $H^\mathrm{sl}$ Majorana spinors}
\label{csspinors}

\subsubsection{Poisson structure}

Turning now to the one-sheeted hyperboloid, recall that the discussion of section \ref{spinortl} resulted in a set of spinors $\ket{z^{1,2}}$ and a mapping to  $H^\mathrm{sl}$ \eqref{tmatch}. In searching for $\mathrm{SU}(1,1)$ representations we again consider a 4-dimensional complex space, but we pick an orthogonal decomposition adapted to the Majorana spinors, i.e. $\mathbb{C}^4 \simeq (\mathbb{C}^2)^1 \oplus (\mathbb{C}^2)^2$ with a choice of basis 
\begin{equation}
 \{\partial_1^1, \partial^1_2\} \oplus \{\partial^2_1, \partial^2_2\}:= \{\partial_1, \overline{\partial}_1\} \oplus \{\partial_2, \overline{\partial}_2\}\,,
\end{equation}
so that the upper index denotes the factor. A general element $u \in (\mathbb{C}^2)^1 \oplus (\mathbb{C}^2)^2$ takes the form $u=u^1+u^2$ for $u^i \in (\mathbb{C}^2)^i$, and there is a linear involution commuting the factors
\begin{equation}
\begin{gathered}
  T: \mathbb{C}^4 \mapsto \mathbb{C}^4 \\
  T\, \partial_i^1 = \partial_i^2\,.
  \end{gathered}
\end{equation}
The first component is equipped with an indefinite hermitian pairing
\begin{equation}
\label{hftl}
\begin{gathered}
  \braket{\cdot,\cdot}_{(\mathbb{C}^2)^1}: \; (\mathbb{C}^2)^1\times (\mathbb{C}^2)^1 \rightarrow \mathbb{C} \\
  (u^1,v^1) \mapsto \braket{u^1,v^1}_{(\mathbb{C}^2)^1}=u^{1\dagger} \sigma_3 v^1\,,
\end{gathered}
\end{equation}
which is invariant under the action of $\mathrm{SU}(1,1)$. Taking inspiration from before, we consider the symplectic form
\begin{align*}
  & i \omega( u^2_i, v^1_j)= \braket{T\overline{u_i^2}, v^1_j}_{(\mathbb{C}^2)^1}\,, \\
  &  \omega(u^1, v^1)= \omega(u^2, v^2)=0\,,
\end{align*}
where antisymmetry requires that $\braket{u^2, v^2}_{(\mathbb{C}^2)^2}=u^{2 \dagger}\sigma_3 v^2$. Note that $\omega$ has the same structure as the symplectic form defined in equation \eqref{symplecticsl}, but now with an additional involution map $T$ (which was complex conjugation itself in the former case). Explicitly it takes the form
\begin{equation}
\label{symplectict}
  \omega=i ( \diff z_1 \wedge \diff \overline{z}_2 -\diff \overline{z}_1 \wedge \diff z_2 )\,,
\end{equation}
leading to the only non-vanishing Poisson brackets\footnote{For the reader's convenience we reiterate that the double-index notation is to be understood as
\begin{equation*}
z^1_1:=z_1\,, \quad z^1_2:= \overline{z}_1\,, \quad z_1^2=z_2\,, \quad z_2^2=\overline{z}_2\,.
\end{equation*}}
\begin{equation}
\label{poissonmajorana}
  \{\overline{z}^1_i, z^2_j\}=i{\sigma_3}_{ij}\,.
\end{equation}

\subsubsection{Lie algebra realization}

The vector components $v^i=\frac{1}{2}\braket{z^2|\sigma_3 \varsigma^i |z^1}$ determined by equation \eqref{tmatch} read
\begin{equation}
  v^1=\frac{1}{2}(z_1 \overline{z}_2 + \overline{z}_1 z_2)\,, \quad v^2=\frac{1}{2}(z_1 z_2 + \overline{z}_1 \overline{z}_2)\,, \quad v^3=\frac{i}{2}(z_1 z_2-\overline{z}_1 \overline{z}_2)\,,
\end{equation}
with norm 
\begin{align}
|v|^2&=(v^1)^2-(v_2)^2-(v_3)^2 \nonumber \\
&=\frac{1}{4}(z_1 \overline{z}_2-\overline{z}_1 z_2)^2 \leq 0\,,
\end{align}
which is non-positive as intended. The component functions satisfy the same Poisson algebra as equation \eqref{algsl},
\begin{equation}
\label{majoranaalgebra}
  \{v^i, v^j\}=-\epsilon^{ijk}\eta_{kl} v^l\,,
\end{equation}
which can now be recognized as being analogous to the $\mathfrak{su}(1,1)$ Lie algebra. Using the symplectic form \eqref{symplectict}, the functions $v^i$ induce the Hamiltonian vector fields
\begin{equation}
  X^1=-\frac{i}{2}\left(z_1 \partial_1 + z_2 \partial_2 -  \overline{z}_1 \overline{\partial}_1 - \overline{z}_2 \overline{\partial}_2 \right)\,,
\end{equation}
\begin{equation}
  X^2=-\frac{i}{2}\left(\overline{z}_1 \partial_1 + \overline{z}_2 \partial_2 - z_1 \overline{\partial}_1 - z_2 \overline{\partial}_2 \right)\,,
\end{equation}
\begin{equation}
  X^3=-\frac{1}{2}\left(\overline{z}_1 \partial_1+\overline{z}_2 \partial_2 +z_1 \overline{\partial}_1 + z_2 \overline{\partial}_2\right)\,,
\end{equation}
with Lie brackets given by $[X^i,X^j]=\epsilon^{ijk}\eta_{kl} X^l$. As we did before, we proceed by projecting onto the first factor $\mathbb{C}^2_1$ and subsequently identifying $(X^1, X^2, X^3) \mapsto (iL^3, iK^1, iK^2)$. One obtains in this manner the generators 
\begin{equation}
\label{gentl}
  L^3=\frac{1}{2}\left(\overline{z}\overline{\partial}-z \partial\right)\,, \quad  K^1=\frac{1}{2}\left(z  \overline{\partial}-\overline{z} \partial \right)\,, \quad K^2=\frac{i}{2}\left(z  \overline{\partial}+\overline{z} \partial \right)\,, 
\end{equation}
having done away with the subscript $z_1 \mapsto z$, since it is for now irrelevant. There is again an immediate outer automorphism induced by the mapping $(z,\overline{z}) \mapsto (\overline{z},z)$,
\begin{equation}
\label{kparity}
  P L^3 P^{-1} = -L^3\,, \quad P K^1 P^{-1} = -K^1\,, \quad P K^2  P^{-1} =K_2\,,
\end{equation}
which is precisely the parity map of equation \eqref{parity}. The Casimir operator in this realization reads
\begin{align}
\label{casimirtl}
  Q&=(L^3)^2-(K^1)^2-(K^2)^2 \nonumber \\
  &=\frac{1}{4} \left[ 2(z \partial+\overline{z} \overline{\partial}) + (z \partial+\overline{z} \overline{\partial}) ^2 \right] \,,
\end{align}
and the ladder operators are given by
\begin{equation}
  L^+=-\overline{z}\partial\,, \quad L^-=z \overline{\partial}\,.
\end{equation}

\subsubsection{Group representations}

It is once more the case that the Lie algebra representation \eqref{gentl} induces a group representation. We take as carrier space the ring of formal power series $\mathbb{C}[[z,\overline{z}]]$, and one can retroactively check that the representation 
\begin{equation}
  \begin{gathered}
  D: \mathrm{SU}(1,1) \rightarrow \mathcal{O}\left(\mathbb{C}[[z,\overline{z}]]\right) \\
  D(g) f(z, \overline{z}) \mapsto  f ( g^{-1}\triangleright  (z, \overline{z}))\,, \quad g\triangleright  (z, \overline{z})= g\begin{pmatrix}
    z \\ \overline{z} \end{pmatrix}\,,
  \end{gathered}
\end{equation} 
does lead to the realization of equation \eqref{gentl}. Regarding the eigenfunctions of the Casimir element \eqref{casimirtl}, note that any solution to 
\begin{equation}
  (z \partial+\overline{z} \overline{\partial})f(z,\overline{z}) = (-2j-2) f(z,\overline{z})
\end{equation}
is also an eigenfunction\footnote{Although one could have chosen to pick $2j$ as the degree of homogeneity as in \cite{Bargmann:1946me}, we choose to match the convention of \cite{Lindblad:1969zz}. The eigenvalue of $Q$ agrees in both cases, and once one fixes $j=-\frac{1}{2}+is$ both conventions are related by complex conjugation.} of $Q$. The differential equation \eqref{euler} of the previous subsection is analogous to the above up to the fact that $z$ and $\overline{z}$ are complex conjugated. One is thus led to consider homogeneous functions as possible solutions, and indeed the homogeneous series $f_j(r z,r \overline{z})=r^{-2j-2} f_j (z, \overline{z}),\, r \in \mathbb{R}$ solve the differential problem. Hence the $Q$ eigenfunctions satisfy
\begin{equation}
  Q f_j = j(j+1) f_j\,. 
\end{equation}
 
\subsubsection{Hermitian inner product and unitarity}

We again consider the Haar measure in terms of two complex variables $z_1,z_2$. Let $f_j(z_1)$ and $g_{j'}(z_1)$ be two homogeneous functions as above, having reinstated the lower index of $z_1$. Then
\begin{align}
\label{preinnertl}
  &\frac{1}{\pi^2} \int  Dz_1 Dz_2 \; \delta(\braket{z|\sigma_3|z}-1) \overline{f}_j(z_1) g_{j'}(z_1)\nonumber \\
  &= \frac{1}{\pi} \int \diff \theta r\diff r \,\,  \Theta(r-1) \overline{f}_j\left(r e^{i \theta} \right) g_{j'}\left(r e^{i \theta} \right) \nonumber \\
  &= \frac{1}{\pi} \int \diff \theta \int_0^\infty \diff k \, e^{2k[(j'+\overline{j})+1]} \overline{f}_j\left(e^{i \theta}\right) g_{j'}\left(e^{i \theta}\right)\,,
\end{align}
where $z_1=:r e^{i\theta}$ and $k:=\ln r$. Equation \eqref{preinnertl} contains an integral representation of the Dirac delta when the exponent of the last line is purely imaginary. Indeed, setting 
\begin{equation}
  j+\frac{1}{2}=is\,, \quad s \in \mathbb{R}^+\,,
\end{equation}
the whole expression reduces to 
\begin{equation}
  \frac{1}{\pi^2} \int  Dz_1 Dz_2 \; \delta(\braket{z|\sigma_3|z}-1) \overline{f}_j(z_1) g_{j'}(z_1) \sim \delta(s-s') \int_{S^1} \diff \theta\,  \overline{f}_j\left(e^{i \theta}\right) g_{j'}\left(e^{i \theta}\right)\,,
\end{equation}
suggesting one ought to define the inner product 
\begin{equation}
\label{preinnersl2}
  \braket{f,g}=\frac{1}{2\pi}\int_{S^1} \overline{f}(z) g(z)
\end{equation}
over the circle. The formal series $\mathbb{C}[[z,\overline{z}]]$ can be restricted to functions on $S^1$ with period $p$, and upon completion through the norm induced by \eqref{preinnersl2} one gets the Hilbert space of square-integrable functions $L^2_p(S^1)$.  

The invariant pairing defined in \eqref{preinnersl2} agrees exactly with the inner product on the continuous series of unitary representations of $\mathrm{SU}(1,1)$ \cite{Bargmann:1946me}, and hence the representations associated to the $H^{\mathrm{sl}}$ Majorana spinors do correspond to the continuous series. The representation $D(g)f(z,\overline{z})=f(g^{-1}(z,\overline {z}))$ matches the multiplier representation defined in  \cite[Ch. VII]{gelfand1966generalized} through the homogeneity property on $\psi(\theta)=f(e^{i\theta})$,
\begin{equation}
  D(g) \psi(\theta)= |\overline{\alpha}e^{i\theta} - \beta e^{-i\theta}|^{-2j-2} \psi\left(\arg \frac{\overline{\alpha}e^{i\theta} - \beta e^{-i\theta}}{|\overline{\alpha}e^{i\theta} - \beta e^{-i\theta}|} \right)\,, \quad g=\begin{pmatrix}
    \alpha & \beta \\ \overline{\beta} & \overline{\alpha}
  \end{pmatrix}\,. 
\end{equation}
With respect to the labels $j=-\frac{1}{2}+is$ the Casimir operator acts as
\begin{equation}
  Q \sim -\left(\frac{1}{4}+s^2\right) <0 \,, 
\end{equation}
and as expected its spectrum is always negative.

\subsubsection{The $L^3$ eigenbasis}

The functions 
\begin{equation}
  f_{j,m}(z)= e^{i\varphi_{j,m}}  z^{-j-1-m} \overline{z}^{-j-1+m}  \quad \left(\; \underset{S^1}{\rightarrow} \; \psi_m(\theta)=e^{i\varphi_{j,m}}  e^{-2i \theta m} \right)
\end{equation}
with $e^{i\varphi_{j,m}}=\left(\frac{\Gamma(m-\overline{j})}{\Gamma(m-j)}\right)^{\frac{1}{2}}$ are homogeneous of degree $-2j-2$, and satisfy
\begin{equation}
  L^3 f_{j,m} = m f_{j,m}\,.
\end{equation} 
The range of the magnetic index can be determined by requiring the representation property $D^j(\mathds{1})=\mathds{1}_j$, i.e.
\begin{equation}
  D^j(e^{4i\pi L^3})f_{j,m}=e^{4i\pi m}f_{j,m}\overset{!}{=}f_{j,m}\,,
\end{equation}
by virtue of which $m$ must take integer and half-integer values. This splits the continuous series into two classes of representations: the \textit{integer series} $\mathcal{C}^0_j$, for which 
\begin{equation}
  m = \mathbb{Z}\,, \quad D(-\mathds{1})\psi_m(\theta)=\psi_m(\theta)\,, \quad \psi(\theta+\pi)=\psi(\theta)\,,
\end{equation}
and the \textit{half-integer} series $\mathcal{C}^{\frac{1}{2}}_j$, characterized by 
\begin{equation}
  m = \frac{1}{2}+\mathbb{Z}\,, \quad D(-\mathds{1})\psi_m(\theta)=-\psi_m(\theta)\,, \quad \psi(\theta+\pi)=-\psi(\theta)\,.
\end{equation}
The states $\psi_m$ in $\mathcal{C}^0_j$ and $\mathcal{C}^{\frac{1}{2}}_j$ thus respectively constitute an orthonormal basis for $L^2_\pi(S^1)$ and $L^2_{2\pi}(S^1)$, as is known from Fourier analysis. The phase $e^{i\varphi_{j,m}}$ was chosen such that the ladder operators have the same expression\footnote{Under $k \mapsto j$. Note that $[(m\pm j\pm 1)(m\mp j)]^{\frac{1}{2}}=|m\mp \overline{j}|$ for the continuous series. The chosen  $e^{i\varphi_{j,m}}$ phase is a solution to the recursion relation $(m-\overline{j})e^{i \varphi_{j,m}}=|m-\overline{j}|e^{i \varphi_{j,m+1}}$.} as in the discrete series, i.e. 
\begin{equation}
  L^\pm f_{j,m} = \sqrt{(m\pm j\pm 1)(m\mp j)} f_{j,m\pm 1}\,.
\end{equation}
The parity operator acts by conjugation $z\mapsto \overline{z}$, so
\begin{equation}
  P f_{j,m}=e^{i \varphi_{j,m}} e^{-i \varphi_{j,-m}} f_{j,-m}=e^{i\pi m} f_{j,-m}\,, 
\end{equation}
confirming that $P^2=\mathds{1}$ is an involution. 
Finally, irreducibility for each $\mathcal{C}^\delta_j$, $\delta=0,\frac{1}{2}$ can be proven following the same strategy as for the discrete case. 

\section{Twistor gauge reduction for Majorana spinors}

Although the discussion of section \ref{su11reps} recovers only well-known results from the representation theory of $\mathrm{SU}(1,1)$ \cite{Bargmann:1946me}, it nonetheless validates the choice of symplectic form \eqref{symplectict} for Majorana spinors. This structure is now put to good use in identifying the space of Majorana spinors with $T^*\mathrm{SU}(1,1)$, thus attributing to the latter an interpretation as the phase space of time-like 2-surfaces in $\mathbb{R}^{1,2}$. 

Following the strategy of \cite{Freidel:2010bw}, we consider a graph link together with two sets of Majorana pairs,
\begin{equation*}
 (\ket{z^1}, \ket{z^2}) \scalebox{0.8}{\begin{tikzpicture}
	\begin{pgfonlayer}{nodelayer}
		\node [style=none] (0) at (0, 0) {};
		\node [style={coh_black}] (2) at (-2.5, 0) {};
		\node [style={coh_black}] (3) at (2.5, 0) {};
		\node [style=none] (4) at (-3.75, 0) {};
		\node [style=none] (5) at (3.75, 0) {};
		\node [style=none] (6) at (-1.25, 0.5) {$>$};
	\end{pgfonlayer}
	\begin{pgfonlayer}{edgelayer}
		\draw [style=line, bend left=330, looseness=1.25] (0.center) to (2);
		\draw [style=line, bend right, looseness=1.25] (0.center) to (3);
	\end{pgfonlayer}
\end{tikzpicture}
} (\ket{w^2}, \ket{w^1})\,.
\end{equation*}
A ``parallel transport'' can be constructed as 
\begin{equation}
  g(z,w)=\frac{\ket{w^1} \bra{z^1} \sigma_3 + \ket{w^2} \bra{z^2} \sigma_3}{\sqrt{\braket{w^2|\sigma_3|w^1}\braket{z^2|\sigma_3|z^1}}}\,,
\end{equation}
taking source spinors to target spinors
\begin{equation}
\label{ptransportt}
  g(z,w)\ket{z^1}=\sqrt{\frac{\braket{z^2|\sigma_3| z^1}}{\braket{w^2|\sigma_3|w^1}}}\, \ket{w^2}\,, \quad   g(z,w)\ket{z^2}=\sqrt{\frac{\braket{z^2|\sigma_3| z^1}}{\braket{w^2|\sigma_3|w^1}}}\, \ket{w^1}\,,   
\end{equation}
and transforming as $g(h_1z, h_2 w)=h_2 g(z,w) h_1^{-1}$ for $h \in \mathrm{SU}(1,1)$. This suggests considering an area-matching constraint
\begin{equation}
  \mathcal{C}=\braket{w^2|\sigma_3|w^1}-\braket{z^2|\sigma_3|z^1} \sim 0\,,
\end{equation}
so-called because each of the bilinears corresponds to the norm of the associated geometrical vector, as per equation \eqref{tmatch}; this constraint naturally generalizes the original one of \cite{Freidel:2010bw}. Upon enforcing the constraint it is straightforward to check that $g(z,w)$ commutes with the real structure and that $\det g(z,w)=1$, from where $g(z,w) \in \mathrm{SU}(1,1)$ in the constraint hypersurface, justifying the term ``parallel transport''. Regarding the Poisson brackets we employ \eqref{poissonmajorana}, and it was already established in \eqref{majoranaalgebra} that then $\{v_z^i, v_z^k\}=-\epsilon^{ijk}\eta_{kl}v_z^l$, where $v^i_z:=\braket{z^2|\sigma_3 \varsigma^i |z^1}$ are functions dual through $\omega$ to a complete set of left-invariant vector fields. Straightforward computations show\footnote{That the Poisson commutator of group elements vanishes can be shown by developing the brackets and making use of the identity
\begin{equation}
\sigma_2=-i \frac{\ket{z^1}\bra{\overline{z}^2} - \ket{z^2}\bra{\overline{z}^1}}{\braket{z^2|\sigma_3|z^1}} = -i \frac{\ket{w^1}\bra{\overline{w}^2} -  \ket{w^2}\bra{\overline{w}^1}}{\braket{w^2|\sigma_3|w^1}}\,.
\end{equation}
}
that on the constraint hypersurface
\begin{equation}
  \{g_{ab}(z,w),g_{cd}(z,w)\}=0\,, \quad \{g_{ab}(z,w), v_z^i\}=-i g_{ab}(z,w)\, \varsigma^i\,;
\end{equation}
$\mathcal{C}$ moreover commutes with $g(z,w)$ and $v_z$. These brackets match the Kirillov-Kostant-Souriau Poisson structure on $T^*\mathrm{SU}(1,1)$ \cite{kirillov2004lectures, Rennert:2016rfp}, and thus one has the symplectomorphism
\begin{equation}
 \left[\, (\mathbb{C}^2)^1 \oplus (\mathbb{C}^2)^2 \setminus \{\braket{z^2|\sigma_3|z^1}=0 \}\,\right] /\!/ \, \mathcal{C} \quad  \simeq \quad T^* \mathrm{SU}(1,1) \setminus |v|=0\,,
\end{equation}
which serves as an $\mathrm{SU}(1,1)$ version of the original findings of \cite{Freidel:2010bw}.

\section{Discussion}

Obtaining a well-behaved spin-foam vertex for space- and time-like polygons is still an open problem; insofar as it is natural to expect that a complete quantum gravity theory ought to make no assumptions on the causal character of the different space-time regions\footnote{At least in the bulk, away from the boundary where one would conceivably be able to constrain the relevant physical states.}, it further remains a fundamental one. The principal obstacle for EPRL-type spin-foams lies in the amplitude for time-like triangles, which seems to be  atypical among all other cases \cite{Conrady:2010kc, Liu:2018gfc, Simao:2021qno}:  it makes use of \textit{generalized} eigenstates of a non-compact $\mathrm{SU}(1,1)$ generator, its integral representation is comparatively complicated, and the critical points of the amplitude seem to not be isolated. A closed-form expression for the semiclassical amplitude has not been achieved, nor is it known whether the amplitude is at all finite.

Since the Conrady-Hnybida generalization of the initial EPRL proposal is a rather natural one, it is conceivable that the aforementioned difficulties associated to the time-like amplitude derive not from the general approach, but rather from the choice of boundary states. It is based on this possibility that the results of the present paper have been sought, in the hope that exploring the spinor correspondence to time-like surfaces might inform a different choice of $\mathrm{SU}(1,1)$ boundary states, leading to a better-behaved vertex amplitude. 

To that end, in this paper we have introduced a new correspondence between Majorana spinors and the one-sheeted space-like hyperboloid $H^{\mathrm{sl}} \subset \mathbb{R}^{1,2}$. The correspondence is based on the identification of a common biquaternionic structure in the critical point equations of the EPRL-CH model, which allowed us to generalize the better-understood mappings between Weyl spinors and $S^2, H^{^\pm} \subset \mathbb{R}^{1,2}$. We have further proposed a particular symplectic structure on $\mathbb{C}^4$, which has not only afforded a Majorana spinorial realization of the continuous series of $\mathrm{SU}(1,1)$ representations, but also a symplectomorphism between the former and the phase space $T^*\mathrm{SU}(1,1)$. This last mapping extends the twisted geometries picture to time-like surfaces \cite{Freidel:2010aq, Freidel:2010bw, Livine:2011gp, Rennert:2016rfp}. Curiously, this establishes a bidirectional characterization of $T^*\mathrm{SU}(1,1)$: it can either be understood as the phase space of time-like surfaces in $\mathbb{R}^{1,2}$ with a given symplectic structure, or as the phase space of $\mathbb{R}^{1,2}$ space-like surfaces with a second, different symplectic form:
\begin{equation*}
\scalebox{1}{\begin{tikzpicture}
	\begin{pgfonlayer}{nodelayer}
		\node [style=none] (0) at (0, 0) {$T^*\mathrm{SU}(1,1)$};
		\node [style=none] (1) at (-2.5, -0.5) {};
		\node [style=none] (2) at (2.5, -0.5) {};
		\node [style=none] (3) at (-3.75, -1.25) {};
		\node [style=none] (4) at (3.75, -1.25) {};
		\node [style=none] (5) at (5, -1.25) {};
		\node [style=none] (6) at (6.75, -1.5) {};
		\node [style=none] (7) at (4, -2) {};
		\node [style=none] (8) at (6, -2.25) {};
		\node [style=none] (9) at (5.5, -1.75) {};
		\node [style=none] (10) at (5.75, -1) {};
		\node [style=none] (11) at (6.75, -0.75) {\small $H^{\mathrm{sl}}$};
		\node [style=none] (12) at (6, -3) {\small $ \omega=i ( \diff z_1 \wedge \diff \overline{z}_2 -\diff \overline{z}_1 \wedge \diff z_2 )$};
		\node [style=none] (13) at (-6.25, -1.25) {};
		\node [style=none] (14) at (-8, -1.5) {};
		\node [style=none] (15) at (-5.25, -2) {};
		\node [style=none] (16) at (-7.25, -2.25) {};
		\node [style=none] (17) at (-6.75, -1.75) {};
		\node [style=none] (18) at (-7, -1) {};
		\node [style=none] (19) at (-8, -0.75) {\small $H^{\pm}$};
		\node [style=none] (20) at (-6.5, -3) {\small $\omega=i(\diff z_1\wedge \diff \overline{z}_1 - \diff z_2 \wedge \diff \overline{z}_2)$};
	\end{pgfonlayer}
	\begin{pgfonlayer}{edgelayer}
		\draw [style=arrow] (1.center) to (3.center);
		\draw [style=arrow] (2.center) to (4.center);
		\draw [style=grayfill] (6.center)
			 to (5.center)
			 to (7.center)
			 to [in=180, out=0] (8.center)
			 to cycle;
		\draw [style=arrow] (4.center) to (2.center);
		\draw [style=arrow] (3.center) to (1.center);
		\draw [style=arrow] (9.center) to (10.center);
		\draw [style=grayfill] (14.center)
			 to (13.center)
			 to (15.center)
			 to [in=0, out=180] (16.center)
			 to cycle;
		\draw [style=arrow] (17.center) to (18.center);
	\end{pgfonlayer}
\end{tikzpicture}
} \,.
\end{equation*}
A number of remarks are due on these results, which now follow. 

First, note that the singular behavior of time-like surfaces manifests itself already at the level of the classical phase space, prior to their inclusion in the spin-foam vertex amplitude. Indeed, time-like surfaces are the only ones among those of interest which seem to be associated with Majorana rather than Weyl spinors. At the level of the Conrady-Hnybida extension, time-like surfaces are also the only ones described by coherent states constructed from a boost generator, with the remaining cases being described by the compact rotation generator common to both $\mathrm{SU}(2)$ and $\mathrm{SU}(1,1)$. A possible connection between these facts is to be explored in future work. 

Secondly, the connection between Majorana spinors and the twistorial description of $T^*\mathrm{SL}(2,\mathbb{C})$ remains to be explored. It is known \cite{Livine:2011vk} that the phase space of surfaces in $\mathbb{R}^3$ can be obtained via symplectic reduction from 2-twistor space by first enforcing an appropriate area constraint, yielding $T^*\mathrm{SL}(2,\mathbb{C})$; the usual simplicity constraints then reduce the latter to $T^*\mathrm{SU}(2)$. It would be interesting to understand how the structures introduced in this paper might arise from such a higher-order construction. 

It must also be pointed out that the structures advocated for here substantially depart from the proposal of  \cite{Rennert:2016rfp}. Although both works are in agreement in the characterization of space-like surfaces, Rennert's spinor states for time-like surfaces are still of the Weyl type (the author's treatment is a unified one for both causal types). We believe this to follow from a possible mistake in equation (3.136) of \cite{Rennert:2016rfp}, which - when accounting for a missing global square - shows that Rennert's spinor states inevitably lead to a positive area spectrum (which ought to be negative for time-like areas). 

Finally, and returning to the initial paragraph of this section, the Majorana spinors here introduced do indeed suggest a possible improvement of the boundary states of the CH extension for time-like triangles. Just how so will be shown in a follow-up paper \cite{prep1}, where it will also be argued that such a modification can be used to define a well-behaved and complete spin-foam coherent vertex amplitude for Lorentzian 3-dimensional quantum gravity with space- and time-like boundaries.

{\centering \noindent\rule{2cm}{0.3pt} \\~\\}

\small \noindent J.D.S. gratefully acknowledges support by the Deutsche Forschungsgemeinschaft (DFG,
German Research Foundation) - Projektnummer/project-number 422809950. The author is indebted to Alexander Jercher for useful comments on the text.  

\bibliographystyle{utphys}
\small
\bibliography{Biq.bib}

\providecommand{\href}[2]{#2}\begingroup\raggedright\begin{thebibliography}{10}

\bibitem{Engle:2007wy}
J.~Engle, E.~Livine, R.~Pereira, and C.~Rovelli, ``{LQG vertex with finite Immirzi parameter},'' \href{http://dx.doi.org/10.1016/j.nuclphysb.2008.02.018}{{\em Nucl. Phys. B} {\bfseries 799} (2008) 136--149}, \href{http://arxiv.org/abs/0711.0146}{{\ttfamily arXiv:0711.0146}}.

\bibitem{Holst:1995pc}
S.~Holst, ``{Barbero's Hamiltonian derived from a generalized Hilbert-Palatini action},'' \href{http://dx.doi.org/10.1103/PhysRevD.53.5966}{{\em Phys. Rev. D} {\bfseries 53} (1996) 5966--5969}, \href{http://arxiv.org/abs/gr-qc/9511026}{{\ttfamily arXiv:gr-qc/9511026}}.

\bibitem{Conrady:2010kc}
F.~Conrady and J.~Hnybida, ``{A spin foam model for general Lorentzian 4-geometries},'' \href{http://dx.doi.org/10.1088/0264-9381/27/18/185011}{{\em Class. Quant. Grav.} {\bfseries 27} (2010) 185011}, \href{http://arxiv.org/abs/1002.1959}{{\ttfamily arXiv:1002.1959}}.

\bibitem{Barrett:2009mw}
J.~W. Barrett, R.~J. Dowdall, W.~J. Fairbairn, F.~Hellmann, and R.~Pereira, ``{Lorentzian spin foam amplitudes: Graphical calculus and asymptotics},'' \href{http://dx.doi.org/10.1088/0264-9381/27/16/165009}{{\em Class. Quant. Grav.} {\bfseries 27} (2010) 165009}, \href{http://arxiv.org/abs/0907.2440}{{\ttfamily arXiv:0907.2440}}.

\bibitem{Kaminski:2017eew}
W.~Kaminski, M.~Kisielowski, and H.~Sahlmann, ``{Asymptotic analysis of the EPRL model with timelike tetrahedra},'' \href{http://dx.doi.org/10.1088/1361-6382/aac6a4}{{\em Class. Quant. Grav.} {\bfseries 35} no.~13, (2018) 135012}, \href{http://arxiv.org/abs/1705.02862}{{\ttfamily arXiv:1705.02862}}.

\bibitem{Liu:2018gfc}
H.~Liu and M.~Han, ``{Asymptotic analysis of spin foam amplitude with timelike triangles},'' \href{http://dx.doi.org/10.1103/PhysRevD.99.084040}{{\em Phys. Rev. D} {\bfseries 99} no.~8, (2019) 084040}, \href{http://arxiv.org/abs/1810.09042}{{\ttfamily arXiv:1810.09042}}.

\bibitem{Simao:2021qno}
J.~D. Sim\~ao and S.~Steinhaus, ``{Asymptotic analysis of spin-foams with timelike faces in a new parametrization},'' \href{http://dx.doi.org/10.1103/PhysRevD.104.126001}{{\em Phys. Rev. D} {\bfseries 104} no.~12, (2021) 126001}, \href{http://arxiv.org/abs/2106.15635}{{\ttfamily arXiv:2106.15635}}.

\bibitem{Livine:2007vk}
E.~R. Livine and S.~Speziale, ``{A New spinfoam vertex for quantum gravity},'' \href{http://dx.doi.org/10.1103/PhysRevD.76.084028}{{\em Phys. Rev. D} {\bfseries 76} (2007) 084028}, \href{http://arxiv.org/abs/0705.0674}{{\ttfamily arXiv:0705.0674}}.

\bibitem{Freidel:2010aq}
L.~Freidel and S.~Speziale, ``{Twisted geometries: A geometric parametrisation of SU(2) phase space},'' \href{http://dx.doi.org/10.1103/PhysRevD.82.084040}{{\em Phys. Rev. D} {\bfseries 82} (2010) 084040}, \href{http://arxiv.org/abs/1001.2748}{{\ttfamily arXiv:1001.2748}}.

\bibitem{Freidel:2010bw}
L.~Freidel and S.~Speziale, ``{From twistors to twisted geometries},'' \href{http://dx.doi.org/10.1103/PhysRevD.82.084041}{{\em Phys. Rev. D} {\bfseries 82} (2010) 084041}, \href{http://arxiv.org/abs/1006.0199}{{\ttfamily arXiv:1006.0199}}.

\bibitem{Livine:2011gp}
E.~R. Livine and J.~Tambornino, ``{Spinor Representation for Loop Quantum Gravity},'' \href{http://dx.doi.org/10.1063/1.3675465}{{\em J. Math. Phys.} {\bfseries 53} (2012) 012503}, \href{http://arxiv.org/abs/1105.3385}{{\ttfamily arXiv:1105.3385}}.

\bibitem{Rennert:2016rfp}
J.~Rennert, ``{Timelike twisted geometries},'' \href{http://dx.doi.org/10.1103/PhysRevD.95.026002}{{\em Phys. Rev. D} {\bfseries 95} no.~2, (2017) 026002}, \href{http://arxiv.org/abs/1611.00441}{{\ttfamily arXiv:1611.00441}}.

\bibitem{Livine:2011vk}
E.~R. Livine, S.~Speziale, and J.~Tambornino, ``{Twistor Networks and Covariant Twisted Geometries},'' \href{http://dx.doi.org/10.1103/PhysRevD.85.064002}{{\em Phys. Rev. D} {\bfseries 85} (2012) 064002}, \href{http://arxiv.org/abs/1108.0369}{{\ttfamily arXiv:1108.0369}}.

\bibitem{Woit:2017vqo}
P.~Woit, \href{http://dx.doi.org/10.1007/978-3-319-64612-1}{{\em {Quantum Theory, Groups and Representations}}}.
\newblock Springer, 2017.

\bibitem{Speziale:2012nu}
S.~Speziale and W.~M. Wieland, ``{The twistorial structure of loop-gravity transition amplitudes},'' \href{http://dx.doi.org/10.1103/PhysRevD.86.124023}{{\em Phys. Rev. D} {\bfseries 86} (2012) 124023}, \href{http://arxiv.org/abs/1207.6348}{{\ttfamily arXiv:1207.6348}}.

\bibitem{Conrady:2010sx}
F.~Conrady and J.~Hnybida, ``{Unitary irreducible representations of SL(2,C) in discrete and continuous SU(1,1) bases},'' \href{http://dx.doi.org/10.1063/1.3533393}{{\em J. Math. Phys.} {\bfseries 52} (2011) 012501}, \href{http://arxiv.org/abs/1007.0937}{{\ttfamily arXiv:1007.0937}}.

\bibitem{ruhl1970lorentz}
W.~Ruhl and W.~R{\"u}hl, {\em The Lorentz Group and Harmonic Analysis}.
\newblock Mathematical physics monograph series. W. A. Benjamin, 1970.

\bibitem{Bargmann:1946me}
V.~Bargmann, ``{Irreducible unitary representations of the Lorentz group},'' \href{http://dx.doi.org/10.2307/1969129}{{\em Annals Math.} {\bfseries 48} (1947) 568--640}.

\bibitem{Lindblad:1969zz}
G.~Lindblad and B.~Nagel, ``{Continuous bases for unitary irreducible representations of $\mathrm{SU}(1, 1)$},'' {\em Ann. De L’I.H.P., Sec. A.} {\bfseries 13} (1970) 27--56.

\bibitem{combescure2012coherent}
M.~Combescure and D.~Robert, {\em Coherent States and Applications in Mathematical Physics}.
\newblock Theoretical and Mathematical Physics. Springer Netherlands, 2012.

\bibitem{gelfand1966generalized}
I.~Gelfand, M.~Graev, and N.~Vilenkin, {\em Generalized Functions: Integral geometry and representation theory}, vol.~5.
\newblock Acad. Press, 1966.

\bibitem{kirillov2004lectures}
A.~Kirillov, {\em Lectures on the Orbit Method}.
\newblock Graduate studies in mathematics. American Mathematical Society, 2004.

\bibitem{prep1}
J.~D. Sim\~ao, ``A new 2+1 coherent spin-foam vertex for quantum gravity.'' \textit{In preparation}.

\end{thebibliography}\endgroup

\end{document}